\documentclass{aa}
\usepackage{graphicx}
\usepackage{bm}
\usepackage{times}
\usepackage{amsmath}
\usepackage{amssymb}
\usepackage{natbib}
\usepackage{color}
\usepackage{mathrsfs}
\usepackage[colorlinks,allcolors=blue]{hyperref}
\usepackage{url}
\usepackage{multirow}
\bibpunct{(}{)}{;}{a}{}{,}
\graphicspath{{./fig/}{./png/}}

\newcommand{\lsim}{\, \, \raisebox{-0.8ex}{$\stackrel{\textstyle <}{\sim}$ }}

\usepackage{graphics}
\usepackage{graphicx}
\usepackage{color}

\definecolor{maroon}{cmyk}{0,0.87,0.68,0.32}

\newcommand{\beq}{\begin{equation}}
\newcommand{\eeq}{\end{equation}}
\newcommand{\beqar}{\begin{eqnarray}}
\newcommand{\eeqar}{\end{eqnarray}}

\usepackage{amsmath}
\usepackage{amssymb}

\begin{document}

\authorrunning{Sotani et al.}
\titlerunning{NS model constrained from high QPOs}

%   \title{Constraint on the neutron star model using high-frequency quasi-periodic oscillations in magnetar \\
%   \blu{KK: Using the high-frequency QPOs of GRB 200415A for constraining the neutron star parameters.}\\
   \title{Neutron star mass-radius constraints using the high-frequency QPOs of GRB 200415A}

   \author{ H. Sotani\inst{1,2}, K. D. Kokkotas\inst{3,4}, and N. Stergioulas\inst{4}          }

   \institute{Astrophysical Big Bang Laboratory, RIKEN, Saitama 351-0198, Japan\\
              email: \href{mailto:sotani@yukawa.kyoto-u.ac.jp}{sotani@yukawa.kyoto-u.ac.jp}
   \and
              Interdisciplinary Theoretical \& Mathematical Science Program (iTHEMS), 
              RIKEN, Saitama 351-0198, Japan
   \and
              Theoretical Astrophysics, IAAT, University of T\"{u}bingen, 72076 T\"{u}bingen, Germany
   \and
              Department of Physics, Aristotle University of Thessaloniki, Thessaloniki 54124, Greece
}

   %\date{Received ? / Accepted ?}
\date{\today}

\abstract{Quasi-periodic oscillations (QPOs) observed in a giant flare of a strongly magnetized neutron star (magnetar), are carrying crucial information for extracting the neutron star properties.
   }%
   {The aim of the study is to constrain the mass and radius of the neutron star model for GRB 200415A, 
   %which radiates high-frequency QPOs, 
   by identifying the observed QPOs with the crustal torsional oscillations together with the experimental constraints on the nuclear matter properties. 
   }%
   {The frequencies of the crustal torsional oscillations are determined by solving the eigenvalue problem with the Cowling approximation, assuming a magnetic field of about $10^{15}$G.}
   {We find that the observed QPOs can be identified with several overtones of crustal oscillations, for carefully selected combinations of the nuclear saturation parameters. Thus, we can inversely constrain the neutron star mass and radius for GRB 200415A by comparing them to the values of nuclear saturation parameters obtained from terrestrial experiments.
   }%
   {We impose further constraints on the neutron star mass and radius while the candidate neutron star models are consistent with the constraints obtained from other available astronomical and experimental observations.
   }%
   \keywords{   asteroseismology -- stars: neutron -- stars: oscillations   }

  \maketitle

%____________________________________________________________

%%%%%%%%%%%%%%%%%%%%%%%%%%%%%%%%%%%%%%%%%%%%%%%%
\section{Introduction}
\label{sec:I}
%%%%%%%%%%%%%%%%%%%%%%%%%%%%%%%%%%%%%%%%%%%%%%%%

Neutron stars are remnants of core-collapse supernovae, in which extreme states of matter are realized. The density inside the star easily exceeds the standard nuclear density ($\rho_0$), while the gravitational and magnetic fields inside/around the neutron star are among the strongest observed anywhere in the Universe  \cite{NS}. These extreme matter conditions cannot be reproduced on Earth, and thus the observations of neutron stars and the associated phenomena remain as the only way to extract information about the state of matter at the most extreme densities. 
An example of this ``indirect'' study of nuclear matter under extreme conditions is the discovery of $2M_\odot$ neutron stars \citep{D10,A13,C20}. Based on these observations, the possibility of so-called {\it soft} equations of state (EOSs) for neutron star matter has been practically abandoned (at least within the framework of Einstein's general relativity theory). In particular, many of the EOSs containing hyperons, which generally have a soft core, have been excluded, leading to the so-called hyperon puzzle (but see also e.g. \cite{2023ApJ...942...55S}).
The detection and analysis of the gravitational wave signal during the binary neutron star merger GW170817 \citep{2017PhRvL.119p1101A,LIGOScientific:2018hze} provided information on the tidal deformability of the merging neutron stars, which can be translated to constraints on the neutron star radius, 
%, i.e., $R_{1.4}\le 13.6$ km 
see \cite{2020GReGr..52..109C,2021GReGr..53...27D} for recent reviews.
%\citep{Annala18}
Furthermore, pulsar observations provide useful constraints on the neutron star properties. For example, the light bending due to the strong gravitational field induced by the compact object, a purely relativistic effect, has been proven extremely useful in constraining the neutron star properties. More specifically, since the pulsar light curve depends on stellar compactness, one can constrain the neutron star mass and radius by carefully analyzing the observed light curve (e.g., \citep{PFC83,LL95,PG2003,PO2014,SM18a}). Recently, the Neutron star Interior Composition Explorer (NICER) on the International Space Station via x-ray observations of neutron stars provided constraints on the parameters of two neutron stars, PSR J0030+0451 \citep{Riley19,Miller19} and PSR J0740+6620 \citep{Riley21,Miller21}.  Moreover, for the first time, at the pre-merger stage preceding the GRB211211,  quasi-periodic oscillations
 (QPOs) at $22^{+3}_{-2}$ Hz and $51^{+2}_{-2}$Hz were detected  \cite{2022arXiv220502186X}, suggesting that these features of the precursor signal may have resulted from the resonant shattering (due to tidal interactions) of one of the star's crust prior to coalescence  leading to the excitation of crustal oscillations \citep{Tsang2012,Tsang2013,Suvorov2022}\footnote{Note, however, that it is challenging to reconcile the relatively short lifetime of a magnetar with a long inspiral time of a binary neutron star system driven only by gravitational wave radiation.}%\citep{2022A&A...664A.177S}. 

The modeling of the stellar oscillation spectrum provides a unique way for studying the internal structure of stars. Especially for neutron stars, the oscillations can be associated with the spectrum of the emitted gravitational waves, and in this way, one can reveal details of their internal structure (gravitational wave asteroseismology) \citep{AK1996,AK1998,SKH2004,SYMT2011,PA2012,DGKK2013,2021FrASS...8..166K}. 
Moreover, gravitational wave asteroseismology has been used for analyzing the gravitational signals in numerical simulations of the formation of proto-neutron stars in core-collapse supernovae, e.g., \cite{MRBV2018,TCPOF19,STT2021}. 

In addition, the analysis of the afterglow spectra in magnetar giant flares (SGR 1900+14 and SGR 1806-20) demonstrated a rich spectrum, the so-called QPOs \citep{SW2005,I2005,SW2006}. The analysis suggested that the QPOs were due to magnetoelastic oscillation modes
\citep{2007MNRAS.377..159L,2007MNRAS.375..261S,SW2009,2011MNRAS.410L..37G,GNHL2011,SNIO2012,2012MNRAS.423..811C,Gabler2012,SNIO2013a,Gabler2013,Gabler2016,Gabler18} which provided another opportunity for constraining the parameters of neutron stars.

Recently, another giant flare, GRB 200415A, was detected in the direction of the NGC 253 galaxy by the Atmosphere-Space Interactions Monitor (ASIM) on the International Space Station on the 15th of April 2020, where several high-frequency QPOs, with varying significance,  have been found at 836, 1444, 2132, and 4250 Hz \citep{CT21}. Unlike the QPOs observed in the previous giant flares,  the QPOs in GRB 200415A were confined only at high frequencies, mainly due to the shortness of the observational interval. 
We note that, although GRB 200415A was originally classified as a type I (short) gamma-ray burst, it should be classified as a giant flare of a magnetar, considering the significant restrictions on the energetics and the position of the burst on the $E_{p,i}-E_{iso}$ and $T_{90,i}-EH$ diagrams, where $E_{p,i}$, $E_{iso}$, $T_{90,i}$, and $EH$ are the position of the maximum in the energy spectrum $\nu F_\nu$ in the source frame, the isotropic equivalent of the total energy emitted in the gamma-ray range, the duration in the source frame, and $EH\equiv (E_{p,i}/100 {\rm keV})/(E_{iso}/10^{51} {\rm erg})^{0.4}$, see 
 \cite{MP20}.

\cite{CT21} discuss two alternative explanations (that have previously been proposed in the literature) for the occurrence of the high-frequency QPOs in GRB 200415A. One possible explanation is that the high-frequency QPOs are due to Alfv\'en waves traveling back and forth between the footpoints of magnetic field lines, relatively close to the magnetar surface. This may be the result of a reconnection event, following the development of instability in the magnetosphere \citep{Mahlmann19}. The second possible explanation is the excitation of crustal oscillation, which is a viable scenario for the QPOs reported for SGR 1900+14 and SGR 1806-20, as discussed above. Due to the very short lifetime of the observed QPOs, \cite{CT21} slightly favors the first explanation, without excluding the second.

Here, without attempting to favor one over the other of the two explanations, we examine the consequences of the second one. That is, we try to answer the question: {\it If the high-frequency QPOs in GRB 200415A are due to crustal oscillation, what constraints can be placed on the neutron star EOS?}

Regarding the nature of the oscillations, one may attempt to associate some of the observed high-frequency QPOs with polar-type oscillations of neutron stars, such as the fundamental ($f$-) and pressure ($p$-) modes. Still, the excitation of such global modes during giant flares is questionable, as they require large density variations. It is thus more natural to associate these oscillations with magnetoelastic or pure crustal torsional oscillations, depending on the magnetic field strength.

We note that the field strength estimated from the rotational period and its time derivative seems to be less than $10^{15}$ G for most of the observed magnetars (e.g., \cite{TZW15}).

Here, we will work under the assumption that the strength of the magnetic field in GRB 200415A is in the second region, i.e. that it is $\lsim 10^{15}$G, so that it has a short damping time, but the frequencies are still close to the values of pure crustal oscillations.
%Even though the magnetic field strongly modifies the \nick{crustal} oscillation frequencies and the general structure of the spectrum itself, \nick{here we will work under the assumption that the neutron star involved in GRB 200415A has a relatively weak magnetic field, so that its influence on the crustal oscillation frequencies can be ignored, to a first approximation}. The main reason is the absence of an exact estimation of the magnetic field strength, and the ``guesstimate'' in \cite{CT21} that the magnetic field is ``around $10^{15}$ G''. At this strength the magnetic field influence on the oscillation frequencies and in the structure of the spectrum is minimal. 
This assumption is discussed in more detail in Appendix A, where we show that for high-frequency torsional oscillations, the shift of the crustal mode frequency caused by a magnetic field with strength $\lsim 10^{15}$G is less than the observational uncertainty of the QPOs in GRB 200415A, as reported in \cite{CT21} ($\sim 10\%$). In the remainder of the paper, we will thus use frequencies of pure crustal torsional oscillations (in the limit of no magnetic field).

In Sec.~\ref{sec:II} we discuss the properties of the crust and their relation to torsional oscillations. In  Sec.~\ref{sec:III} we try to identify the observed QPOs with the higher overtones of torsional oscillations to arrive at EOS constraints. The discussion of the results is presented in  Sec. \ref{sec:IV}.

%%%%%%%%%%%%%%%%%%%%%%%%%%%%%%%%%%%%%%%%%%%%%%%%
\section{Crust equilibrium and torsional oscillations}
\label{sec:II}
%%%%%%%%%%%%%%%%%%%%%%%%%%%%%%%%%%%%%%%%%%%%%%%%

Matter in the neutron star crust forms a Coulomb lattice, which behaves as a solid (or liquid crystal). As a result, under perturbations the torsional oscillations are favored. Since the core (inner or outer) behaves as a fluid, the torsional oscillations are confined only in the crust, in the absence of rotation.
Thus for a given EOS describing the core and the crust and for a given central density, one can construct a neutron star model by integrating the Tolman-Oppenheimer-Volkoff (TOV) equations starting with appropriate initial conditions at the center. Here, we take a different approach, specifying only the crust EOS and integrating the TOV equations inwards (starting at the star’s surface), for various sets of mass $M$ and radius $R$ (see \cite{SNIO2013a} for details). 
With this approach, one has to select two boundary conditions ($M$ and $R$), but one can avoid the uncertainty in the core EOS.

For the present study, we adopt the same phenomenological family of crust EOS constructed by \cite{OI2003,OI2007}, the so-called OI-EOSs. 
For this family of EOSs the bulk energy per nucleon, for zero-temperature uniform nuclear matter,  is expressed in the vicinity of the saturation density for the symmetric nuclear matter, $n_0$,  as a function of the baryon number density, $n_{\rm b}$, and the asymmetry parameter, $\alpha$, as
%%%%%%
\begin{equation}
   w = w_0 + \frac{K_0}{18n_0^2}(n_{\rm b}-n_0)^2 
      + \left[S_0 + \frac{L}{3n_0}(n_{\rm b}-n_0)\right]\alpha^2, \label{eq:w}
\end{equation}
%%%%%
where $n_{\rm b}$ and $\alpha$ are defined as $n_{\rm b}=n_{\rm n} + n_{\rm p}$ and $\alpha=(n_{\rm n}-n_{\rm p})/n_{\rm b}$.  Here, $n_{\rm n}$ and  $n_{\rm p}$ are the neutron and the proton number density, respectively. 
%Here $n_{\rm n}$ is the neutron number density and  $n_{\rm p}$ the proton number density. 
The five coefficients $w_0$, $n_0$, $S_0$, $K_0$ and $L$  are the nuclear saturation parameters defining uniquely the  EOS of the crust.  
The first three parameters ($w_0$, $n_0$, and $S_0$) are well constrained from terrestrial experiments \citep{Oertel17,Li19}, while the remaining two parameters ($K_0$ and $L$) can be hardly constrained experimentally. Still, there is progress in this direction and already certain constraints can be set, predicting the following values: %$K_0$ and $L$ are gradually constrained via experiments, whose fiducial value are in the range of 
$K_0=240\pm 20$ MeV \citep{K0} and $L=60\pm 20$ MeV \citep{Li19}.
In fact, the magnetar QPOs in SGR 1806-20 and SGR 1900+14, if identified as crustal torsional oscillations, provide some more stringent constraints to $L$, that is $L_{\rm QPO}=58-73$ MeV  \citep{SIO2018}.  The OI-EOSs are directly characterized by the saturation parameters, but they describe the whole crust region.
%
%%%%%%%%%%%%%%%%%%%%%%%%%%%%%%%%%%%
% Table 1
%%%%%%%%%%%%%%%%%%%%%%%%%%%%%%%%%%%
\begin{table*}
\centering
\caption{The EOS parameters adopted in this study, where $\eta$ and $\varsigma$ are the combination of $K_0$ and $L$ given by $\eta=(K_0 L^2)^{1/3}$ and $\varsigma =(K_0^4 L^5)^{1/9}$. The corresponding transition density from spherical to cylindrical nuclei (SP--C) and that from cylindrical to slab-like nuclei (C--S) are also listed. The asterisk at the value of $K_0$ denotes the EOS model where cylindrical nuclei directly change to uniform matter.
In the rightmost column we also show the relative deviation of the frequencies estimated with the fit given by Eq. (\ref{eq:fitting}) from the eigenfrequencies for the model shown in Fig. \ref{fig:M16R121st}.} 
\begin{tabular}{cccccccc}
\hline\hline
% & $K_0$ (MeV)  & $L$ (MeV) & $-y$ (MeV fm$^3$) & $\eta$ (MeV) & $\varsigma$ (MeV) & SP--C (fm$^{-3}$) & C--S (fm$^{-3}$) & \\
 & $K_0$ (MeV)  & $L$ (MeV) & $\eta$ (MeV) & $\varsigma$ (MeV) & SP--C (fm$^{-3}$) & C--S (fm$^{-3}$) & $\Delta f/f$ (\%)\\
\hline
% & 180 & 5.7   & $1800$ & & 0.06000 & 0.08665 &  \\
 & 180 & 31.0 &   $55.7$ & $67.8$ & 0.05887 & 0.07629 & -1.25 \\
 & 180 & 52.2 &   $78.9$ & $90.5$ & 0.06000 & 0.07186 & 1.14 \\
% & 230 & 7.6   & & & 0.05816 & 0.08355 &  \\
 & 230 & 42.6 &   $74.7$ & $90.1$ & 0.06238 & 0.07671 & -0.12 \\
 & 230 & 73.4 &   $107$ & $122$ & 0.06421 & 0.07099 & 0.81 \\
 & 280 & 54.9 &   $94.5$ & $113$ & 0.06638 & 0.07743 & -1.10 \\
 & 280$^*$ & 97.5 & $139$ & $156$  & 0.06678 & 0.06887 & 0.43 \\
 & 360 & 12.8 &  $38.9$ & $56.4$ & 0.05777 & 0.08217 & 0.62 \\
 & 360 & 76.4 &  $128$ & $152$ & 0.07239 & 0.07797 &  -0.60 \\
% & 360 & 146.1 &   $220$ & & & 0.066 & 0.066 &  \\
\hline\hline
\end{tabular}
\label{tab:EOS}
\end{table*}
%%%%%%%%%%%%%%%%%%%%%%%%%%%%%%%%%%%

The OI-EOSs \citep{OI2007} 
adopted in this study are constructed in such a way that $w_0$, $n_0$, and $S_0$ are optimized for given values of $K_0$ and $L$ to reproduce the experimental data for masses and charge radii of stable nuclei, based on the extended Thomas-Fermi theory.
The EOS parameters adopted here are listed in Table \ref{tab:EOS}. Therefore, for constructing the crust equilibrium model, one needs to select two parameters for the EOS, i.e., $K_0$ and $L$, apart from the global parameters defining the star i.e., the mass ($M$) and radius ($R$).

Torsional oscillations are characterized by the shear modulus, $\mu$, since their frequency scales roughly as  $f	\propto \left(\mu/\rho\right)^{1/2}$ where %$R$ is the radius of the star and 
$\rho$ the local value of the density. 
%\mar{
For the phase composed of spherical nuclei, the shear modulus is formulated as a function of the \emph{ion number density}, the \emph{charge number of nuclei}, and the \emph{Wigner-Seitz cell radius} \citep{OI1990,SHOII1991}. 
When the phase is composed of non-spherical ({cylindrical and} slab-like) nuclei, as proposed by \cite{PP1998}, it behaves more like a liquid crystal. Then the shear modulus {in the phase of slab-like nuclei} ceases to exist and at this part of the crust torsional oscillations are not excited at least at a linear level. 
Thus, torsional oscillations can be excited in two distinct areas of the crust a) in the parts where the phase is composed of spherical and cylindrical nuclei, and b) in the part where the phase is composed of cylindrical holes (tubes) and spherical holes (bubbles) nuclei. This leads to double-layer torsional oscillations in the crust, which can be visualized as a ``lasagna sandwich". With such a structure, all QPOs observed in SGR 1806-20 and SGR 1900+14 can be identified, see \cite{SIO2019}.
%}

In this study, we focus on the newly observed  QPOs in GRB 200415A. These QPOs are of relatively higher frequency, and we suggest that they correspond to overtones of crustal torsional oscillations. Thus, we systematically examine the overtones of the torsional oscillations excited in a phase composed of spherical and cylindrical nuclei. In addition, since some of the neutrons are not bound in the nuclei inside the inner crust, one may take into account the effect of such unbound neutrons on the torsional oscillations. In practice, the ratio of the superfluid to the dripped neutron, $N_{\rm s}/N_{\rm d}$, in the phase composed of spherical nuclei, is calculated by the band theory \citep{Chamel12}. The corresponding ratio in the phase composed of cylindrical nuclei is still unclear. So, in this study, we adopt the approach of \cite{Chamel12} for spherical nuclei and  $N_{\rm s}/N_{\rm d}$ as a parameter for cylindrical nuclei, as in  \cite{SIO2018}.

The governing equation for torsional oscillations is derived from the linearized equations of motions \citep{ST83}, where we adopt the relativistic Cowling approximation as in \cite{2007MNRAS.375..261S}. 
With the Cowling approximation, one neglects the metric perturbations, and thus no gravitational waves are emitted. In the limit of slow rotation, axial oscillations do not induce density perturbations and only emit gravitational waves through very weak current-multipoles. The Cowling approximation is thus a good approximation of computing axial modes in the nonrotating limit.
Then, the appropriate boundary conditions are imposed at the stellar surface and the bottom of the crust (composed of cylindrical nuclei). In this way, a well-defined eigenvalue problem is set for the frequency, $f$, of torsional oscillations. For further details, we refer to  \cite{SNIO2012,SIO2018}.

%%%%%%%%%%%%%%%%%%%%%%%%%%%%%%%%%%%
% Figure 1
%%%%%%%%%%%%%%%%%%%%%%%%%%%%%%%%%%%
\begin{figure}
\begin{center}
\includegraphics[scale=0.6]{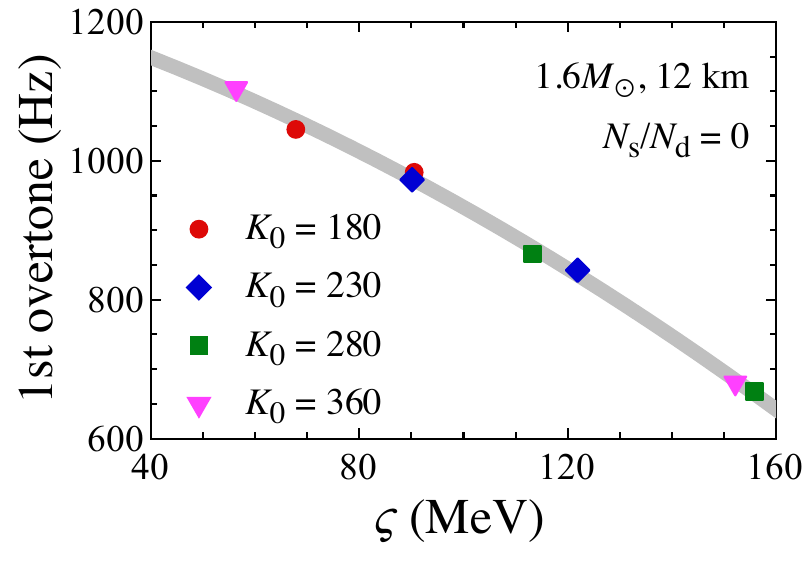} 
\end{center}
\caption{%%
The 1st overtone of the $\ell=2$ crustal torsional oscillations for a neutron star model with $M=1.6M_\odot$, $R=12$km, and $N_{\rm s}/N_{\rm d}=0$. The marks denote the torsional oscillation frequencies estimated for the neutron star models constructed based on the crustal EOSs listed in Table \ref{tab:EOS}, as a function of $\varsigma$. The thick-solid line stands for the quadratic  fitting  given by Eq. (\ref{eq:fitting}). 
}%%
\label{fig:M16R121st}
\end{figure}
%%%%%%%%%%%%%%%%%%%%%%%%%%%%%%%%%%%
The fundamental frequencies of crustal torsional oscillations scale roughly as $f\sim v_s/R$, whereas the overtones scale as $f\sim v_s/\Delta R$, where $v_s\sim \sqrt{\mu/\rho}$ is the shear velocity, and $\Delta R$ the crust thickness  \citep{HC80}. Since the crust thickness depends on both the stellar compactness ($\Delta R/R\sim R/M$) and on the EOS parameters, as shown in \cite{SIO2017b}, the overtones generally depend on $K_0$ and $L$. %for  model with specific mass and radius. 
Actually, in \cite{SIO2018} it was shown that the 1st overtone of crustal torsional oscillations can be described as a function of the parameter
%%%
\begin{equation}
  \varsigma \equiv (K_0^4L^5)^{1/9}. 
  \label{eq:ss} 
\end{equation}
%%%

In fact, as it is shown in Fig. \ref{fig:M16R121st},  the 1st overtone of a neutron star model with $M=1.6M_\odot$, $R=12$ km, and $N_{\rm s}/N_{\rm d}=0$ is expressed as a function of $\varsigma$. The colored marks correspond to the eigenfrequencies, for some typical values of $K_0$, %calculated via the eigenvalue problem, 
while the thick-solid line represents the following  quadratic fitting 
%%%%%%%
\begin{equation}
  {}_\ell t_n = d_{\ell n}^{(0)} + d_{\ell n}^{(1)}\varsigma_{100} + d_{\ell n}^{(2)}\varsigma^2_{100} \, . \label{eq:fitting} 
\end{equation}
%%%%%%
Here,  $\varsigma_{100}\equiv \varsigma/(100 {\rm MeV})$, while the fitting coefficients $d_{\ell n}$, depend on the values of the azimuthal quantum number $\ell$, the nodal number in the corresponding  eigenfunction $n$, the stellar mass $M$, the radius $R$, and the ratio $N_{\rm s}/N_{\rm d}$. Actually, in this study, we further confirm that not only the 1st overtone but also the $n$-th overtones can be fitted by an equation of the form of Eq. (\ref{eq:fitting}). 
For instance, the relative deviation of the frequencies estimated with Eq. (\ref{eq:fitting}) from the eigenfrequencies for the stellar model shown in Fig. \ref{fig:M16R121st}, is listed in the rightmost column in Table \ref{tab:EOS}. That is, the fit given by Eq. (\ref{eq:fitting}) can predict the frequency with less than a few \% accuracy. For the stellar model with $1.4M_\odot$, the coefficients in Eq. (\ref{eq:fitting}) with $(\ell,n)=(2,1)$, $(2,4)$, and $(2,10)$ modes are listed in Table \ref{tab:coefficients}.  
Thus, hereafter, we will discuss the EOS dependence of the overtones using the fitting formula given by Eq. (\ref{eq:fitting}) as a function of $\varsigma$. As discussed earlier, the overtone frequencies generally increase with the stellar compactness, $M/R$ (since $\Delta R/R\sim R/M$) % \citep{SIO2017b} 
and $f\sim v_s/\Delta R$ for the overtones \citep{HC80}.

%%%%%%%%%%%%%%%%%%%%%%%%%%%%%%%%%%%
% Table 2
%%%%%%%%%%%%%%%%%%%%%%%%%%%%%%%%%%%
\begin{table}
\centering
\caption{Coefficients in Eq. (\ref{eq:fitting}) with $(\ell,n)=(2,1)$, $(2,4)$, and $(2,10)$ modes for the neutron star model with $1.4M_\odot$.}
\begin{tabular}{ccccccc}
\hline\hline
 & $n$  & $R$ (km) & $d_0$ & $d_1$ & $d_2$ & \\
\hline 
& $1$ & $10$ & 1566.9 & -266.71 & -131.35 & \\
&  &    $12$ & 1109.7 & -180.73 & -96.521 & \\
&  &    $14$ & 831.40 & -130.70 & -74.339 & \\
& $4$ & $10$ & 3420.1 & -524.71 & -119.01 & \\
&  &    $12$ & 2374.0 & -294.83 & -114.83 & \\
&  &    $14$ & 1745.2 & -172.69 & -104.21 & \\
& $10$ & $10$ & 7589.0 & -2012.7 & 174.84 & \\
&      & $12$ & 5359.0 & -1407.6 & 119.42 & \\
&      & $14$ & 3995.7 & -1031.2 &  81.227 & \\
\hline\hline
\end{tabular}
\label{tab:coefficients}
\end{table}
%%%%%%%%%%%%%%%%%%%%%%%%%%%%%%%%%%%

%%%%%%%%%%%%%%%%%%%%%%%%%%%%%%%%%%%%%%%%%%%%%%%%
\section{Comparison of the overtones with the observed QPOs}
\label{sec:III}
%%%%%%%%%%%%%%%%%%%%%%%%%%%%%%%%%%%%%%%%%%%%%%%%

In this section, we make an attempt to identify which overtones of crustal torsional oscillations match the QPO frequencies observed in GRB 200415A keeping in mind that not all four QPOs were of the same significance. The lower two  were less prominent, which apart from underlying physical reasons (which are unclear, due to the lack of knowledge of the excitation mechanism) should be related also to the short duration of the signal  ($\sim 3$ms).   

\begin{figure}
\begin{center}
\includegraphics[scale=0.6]{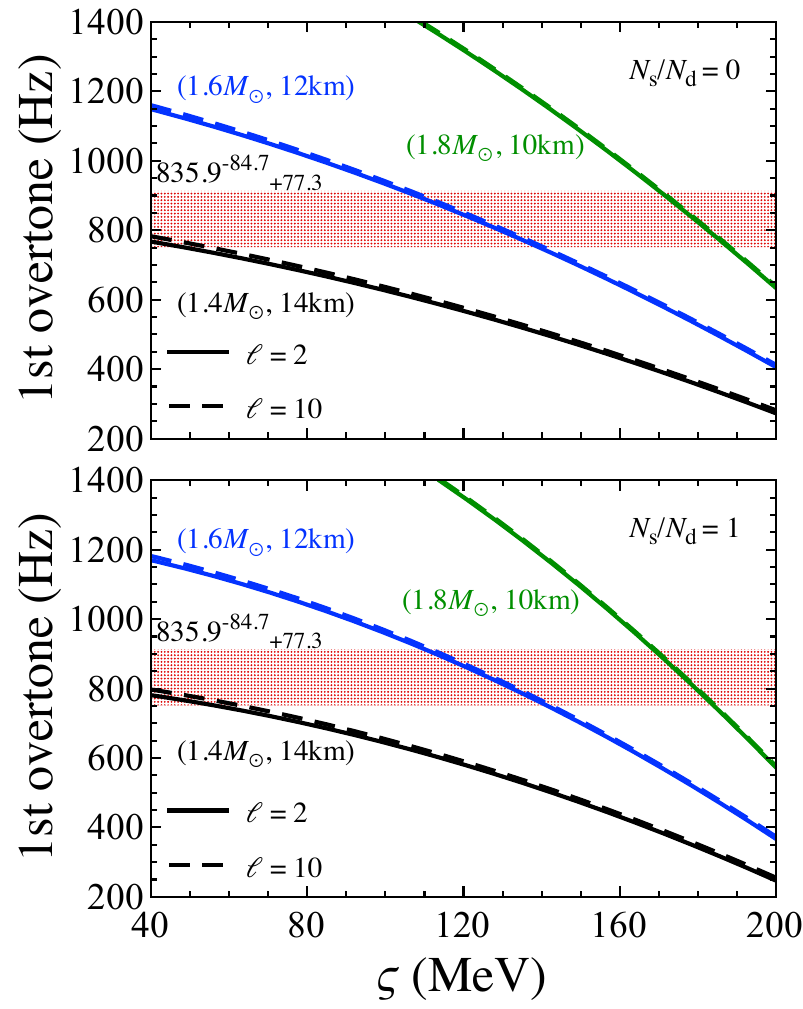} 
\end{center}
\caption{%%
The 1st overtones of the crustal torsional oscillations with $\ell=2$ (solid line) and $\ell=10$ (dashed line) are shown as function of $\varsigma$ given by Eq. (\ref{eq:ss}) for three neutron star models with $(M,R,M/R)=(1.4M_\odot,14{\rm km}, 0.146)$, $(1.6M_\odot,12{\rm km}, 0.197)$ and $(1.8M_\odot,10{\rm km}, 0.266)$. The top panel corresponds to $N_{\rm s}/N_{\rm d}=0$  while the bottom panel to $N_{\rm s}/N_{\rm d}=1$.  In both panels, for reference, we draw (with a shaded thick line) the lowest QPO frequency found in GRB 200415A, i.e., 835.9$^{-84.7}_{+77.3}$ Hz \citep{CT21}. 
}%%
\label{fig:1st}
\end{figure}

As a first step, we will consider the correspondence between the 1st overtone and the lowest QPO frequency extracted from GRB 200415A, i.e., 835.9$^{-84.7}_{+77.3}$ Hz \citep{CT21}. 
In Fig. \ref{fig:1st}, we show the 1st overtones for three characteristic neutron star models with $(M,R)=(1.4M_\odot, 14 {\rm km})$, $(1.6M_\odot, 12 {\rm km})$, and $(1.8M_\odot, 10 {\rm km})$. The solid and dotted lines denote the frequencies with $\ell=2$ and $\ell=10$, while the top and bottom panels correspond to the results with $N_{\rm s}/N_{\rm d}=0$ and $N_{\rm s}/N_{\rm d}=1$, respectively. From this figure, it is obvious that the dependencies of the overtones on $\ell$ and $N_{\rm s}/N_{\rm d}$ is very weak, as mentioned in  \cite{HC80,SIO2018}. So, hereafter, we will only consider the $\ell=2$ overtones for the neutron star model with $N_{\rm s}/N_{\rm d}=0$. In addition, it is noticeable that the 1st overtones increase with compactness as mentioned in the previous section. Actually,  for the three models considered here, $(M,R)=(1.4M_\odot, 14 {\rm km})$, $(1.6M_\odot, 12 {\rm km})$ and $(1.8M_\odot, 10 {\rm km})$, the compactness is $M/R=0.148$, $0.197$ and $0.266$, correspondingly. 
As a result, for identifying the 836 Hz QPO as the 1st overtone, the suitable values of $\varsigma$ should increase as the neutron star compactness becomes higher.
%%%%%%%%%%%%%%%%%%%%%%%%%%%%%%%%%%%
% Figure 2
%%%%%%%%%%%%%%%%%%%%%%%%%%%%%%%%%%%
%%%%%%%%%%%%%%%%%%%%%%%%%%%%%%%%%%%

%%%%%%%%%%%%%%%%%%%%%%%%%%%%%%%%%%%
% Figure 3
%%%%%%%%%%%%%%%%%%%%%%%%%%%%%%%%%%%
\begin{figure}
\begin{center}
\includegraphics[scale=0.6]{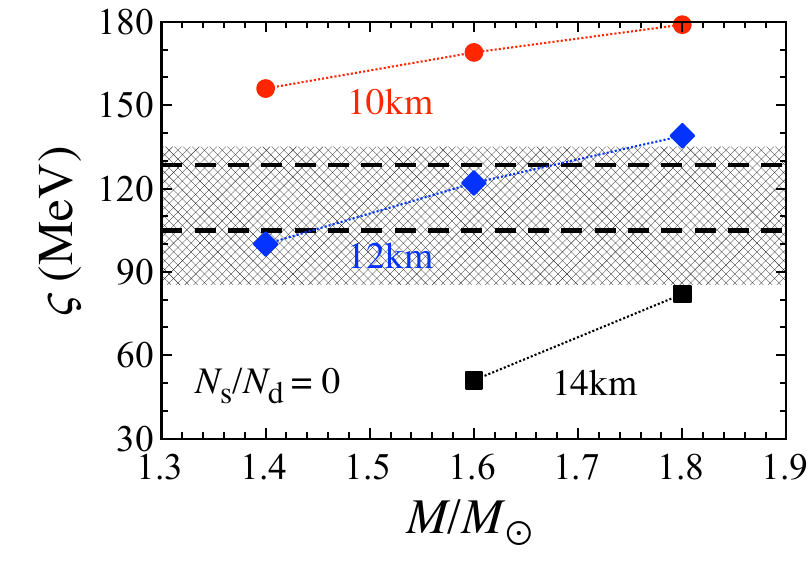} 
\end{center}
\caption{%%
Suitable values of $\varsigma$ for identifying the 835 Hz QPO as the 1st overtone for various stellar masses and radii with $N_{\rm s}/N_{\rm d}=0$ are shown. The shaded region corresponds to the range of $\varsigma=85.3-135.1$MeV, estimated for the fiducial values  $L=60\pm 20$ and $K_0=240\pm 20$ MeV. The dashed lines define the range of $\varsigma_{\rm QPO} = 104.9-128.4$ MeV, corresponding to  $L_{\rm QPO}=58-73$ MeV.
}%%
\label{fig:sigma-1st}
\end{figure}
%%%%%%%%%%%%%%%%%%%%%%%%%%%%%%%%%%%

In Fig. \ref{fig:sigma-1st}, we show the suitable value of $\varsigma$ for identifying the 836 Hz QPO as the 1st overtone of crustal torsional oscillations for some typical neutron star models. 
More specifically,  the shaded region corresponds to neutron star models for  $\varsigma = 85.3 - 135.1$ MeV,  corresponding to
%with the fiducial value of $K_0$ and $L$ i.e., 
$K_0=240\pm 20$ MeV and $L=60\pm 20$ MeV. In addition,  we draw the dashed lines at $\varsigma =104.9$ and $\varsigma = 128.4$ MeV, defining the estimated range for the tighter range $L_{\rm QPO}=58-73$ MeV. The latter bounds have been set once the magnetar QPOs in SGR 1806-20 and 1900+14 were identified as crustal torsional oscillations \citep{SIO2018}. %, using the fiducial value of $K_0$. 
In this figure, it is noticeable that the suitable value of $\varsigma$, for identifying the 836 Hz QPO  as the 1st overtone, increases with $M$, for fixed neutron star radius, and that some characteristic neutron star models,  e.g., with masses $M=1.4M_\odot$, $1.6M_\odot$, $1.8M_\odot$ and  $R=10$ km, do not fit with the observational data.

\begin{figure}
\begin{center}
\includegraphics[scale=0.6]{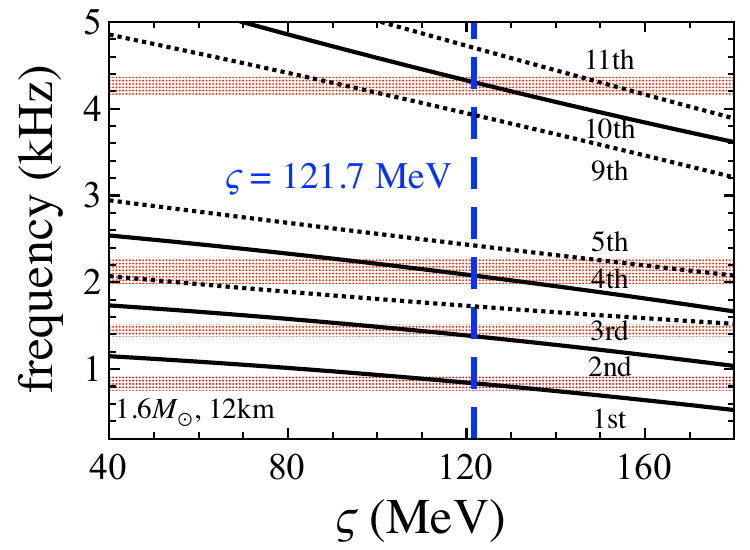} 
\end{center}
\caption{%%
Correspondence of the observed QPOs to the overtones for the neutron star model with $M=1.6M_\odot$ and $12$ km. The QPOs observed in GRB 200415A can be identified with the 1st, 2nd, 4th, and 10th overtones of crustal torsional oscillations, which leads to the suitable value of $\varsigma$ for the adopted stellar model as $\varsigma = 121.7$ MeV.}
\label{fig:4QPOs-M16R12}
\end{figure}

The next step is to associate all four QPO frequencies observed in GRB 200415A with overtones of the crustal torsional oscillations. In fact, we find that it is feasible to identify all four observed QPOs as the 1st, 2nd, 4th, and 10th overtones for neutron stars models with certain values of mass and radius. In Fig. \ref{fig:4QPOs-M16R12}, we show the suggested identification for one of the three neutron star models considered earlier i.e., the one with $M=1.6M_\odot$ and $R=12$ km. In this figure, the correspondence for $\varsigma = 121.7$ MeV for the aforementioned neutron star model is apparent. Since the overtones vary as they depend on the neutron star mass and radius, the suitable value of $\varsigma$ for identifying the observed QPO frequencies varies accordingly. In principle, one could identify all observed QPOs with other combinations of the overtones, but it would not be  consistent with the saturation parameters constrained via experiments (see Appendix \ref{sec:appendix_2} for details).

%%%%%%%%%%%%%%%%%%%%%%%%%%%%%%%%%%%
% Figure 4
%%%%%%%%%%%%%%%%%%%%%%%%%%%%%%%%%%%
%%%%%%%%%%%%%%%%%%%%%%%%%%%%%%%%%%%

\begin{figure}
\begin{center}
\includegraphics[scale=0.6]{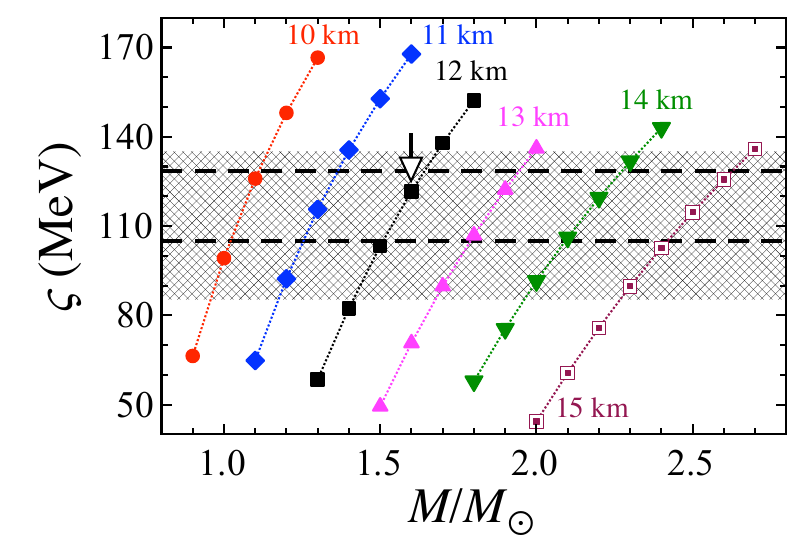} 
\end{center}
\caption{%%
The suitable value of $\varsigma$ (colored marks) for simultaneous identification of the four QPOs observed in GRB 200415A as the 1st, 2nd, 4th, and 10th overtones of the crustal torsional oscillations for various neutron star models. The meaning of the shaded region and dashed lines is the same as in Fig. \ref{fig:sigma-1st}.
The arrow denotes the particular model discussed in Fig. 4, where the extracted value is $\varsigma = 121.7$ MeV.
}%%
\label{fig:sigma-4QPOs}
\end{figure}

In Fig. \ref{fig:sigma-4QPOs}, we show (by colored marks) the extracted values of $\varsigma$ for a wider range of neutron star models. In all of them, we identified the observed QPO frequencies as the 1st, 2nd, 4th, and 10th overtones. %frequencies. 
As in Fig.~\ref{fig:sigma-1st}, in this figure we draw the fiducial range of $\varsigma$ (shaded region) and the range for $\varsigma_{\rm QPO}$ (dashed lines) as constrained by QPO observations.
We note that for the stellar models, 
{which are not considered in Fig.~\ref{fig:sigma-4QPOs}}, 
e.g., for a neutron star model with $M=1.8M_\odot$ and 10 km, the QPO frequencies cannot be identified with the same set of overtones.
%frequencies \blu{(the 1st, 2nd, 4th, and 10th)}. 
%In any case, for the stellar models shown in Fig. \ref{fig:sigma-4QPOs}, one can identify the QPO frequencies  observed in GRB 200415A as the 1st, 2nd, 4th, and 10th overtones.

%%%%%%%%%%%%%%%%%%%%%%%%%%%%%%%%%%%
% Figure 5
%%%%%%%%%%%%%%%%%%%%%%%%%%%%%%%%%%%
%%%%%%%%%%%%%%%%%%%%%%%%%%%%%%%%%%%

\begin{figure}
\begin{center}
\includegraphics[scale=0.6]{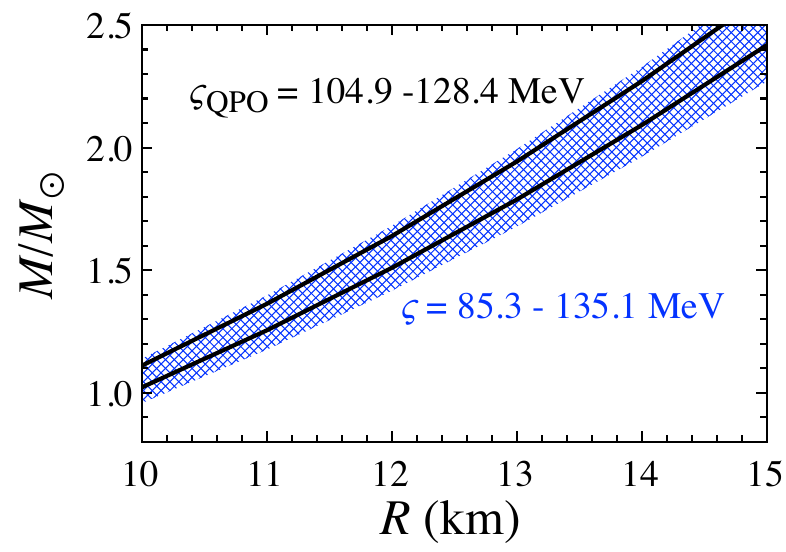} 
\end{center}
\caption{%%
Constraints on neutron star mass and radius set by the simultaneous identification of all four observed QPOs in GRB 200415A as the 1st, 2nd, 4th, and 10th overtones of the crustal torsional oscillations. Here we accepted only the models meeting the constraints $\varsigma=85.3-135.1$ MeV (shaded region) and $\varsigma_{\rm QPO} = 104.9-128.4$ MeV (bounded by solid lines).
}%%
\label{fig:MR-sigma}
\end{figure}

Fig. \ref{fig:MR-sigma} is a mass vs radius diagram on which we extract the constraints set in Fig. \ref{fig:sigma-4QPOs}
for the mass, radius and $\varsigma$. The shaded region (the region enclosed with the solid lines) corresponds to the mass vs radius constraints obtained for $\varsigma=85.3-135.1$ MeV ($\varsigma_{\rm QPO}=104.9-128.4$ MeV).

%%%%%%%%%%%%%%%%%%%%%%%%%%%%%%%%%%%
% Figure 6
%%%%%%%%%%%%%%%%%%%%%%%%%%%%%%%%%%%
%%%%%%%%%%%%%%%%%%%%%%%%%%%%%%%%%%%

\begin{figure}
\begin{center}
\includegraphics[scale=0.6]{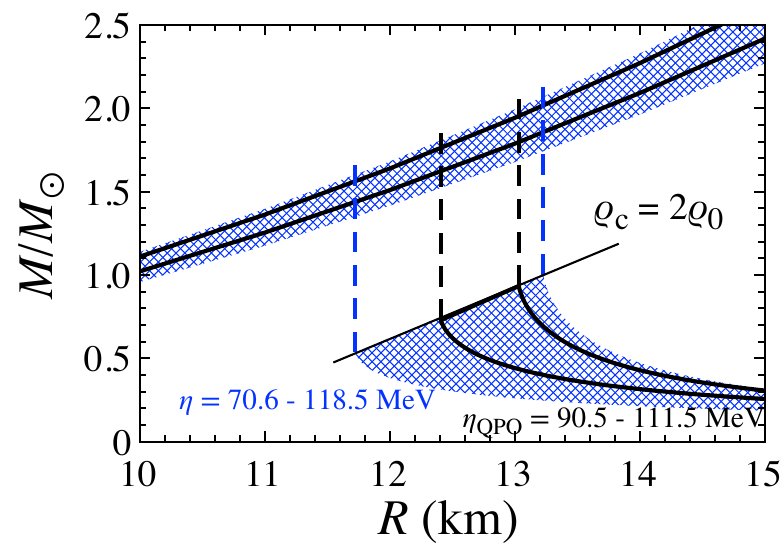} 
\end{center}
\caption{%%
The allowed range for the neutron star mass and radius obtained from the QPOs observed in GRB 200415A, as shown in Fig. \ref{fig:MR-sigma}, is compared to the mass and radius range derived via the empirical formula derived in \citet{SIOO14} (the right-bottom region).  
For defining the low mass models we used  $\eta = 70.6-118.5$ MeV (shaded region) and $\eta_{\rm QPO} = 90.5-111.5$ MeV (the region enclosed by the solid lines). Here we made the assumption that $\rho_c \le 2\rho_0$.
The (black and blue) dashed lines denote the neutron star radii expected from the empirical relation
for $\eta=70.6$, 90.5, 111.5, and 118.5 MeV with $\rho_c=2\rho_0$.
}
\label{fig:MR-QPO}
\end{figure}

It has been shown, that one can estimate the mass and radius of low-mass neutron stars, using another combination of $L$ and $K_0$, i.e., $\eta=(K_0 L^2)^{1/3}$ \citep{SIOO14,SNN22}\footnote{A similar discussion is also possible using another combination of the nuclear saturation parameter given by $\eta_\tau=(-K_\tau L^5)^{1/6}$, where $K_\tau$ is the isospin dependence of incompressibility for asymmetric nuclear matter \citep{SO22}, or directly using the experimental observables, such as neutron skin thickness or dipole polarizability for neutron-rich nuclei \citep{SN23}}. According to this approach, for neutron star models with central density,  $\rho_{\rm c}\le 2\rho_0$,
its mass and gravitational redshift can be expressed as a function of $\eta$ and the normalized $\rho_{\rm c}$. This leads to a relation between the neutron star mass and radius, once $\eta$ is fixed. 
Then we adopt a range for $\eta= 70.6-118.5$ MeV, corresponding 
to  $\varsigma=85.3-135.1$ MeV and $\eta_{\rm QPO}=90.5-111.5$ MeV, 
corresponding to $\varsigma_{\rm QPO} =104.9-128.4$ MeV. The expected region of the neutron star mass and radius with $\rho_{\rm c}\le 2\rho_0$, is given by the shaded region and the region enclosed by the solid lines, on the bottom-right side in Fig. \ref{fig:MR-QPO}. If the resultant region intersects with the region shown in Fig. \ref{fig:MR-sigma}, one could set further constraints to the parameters of the neutron star model corresponding to GRB 200415A. 
Unfortunately, the shaded area corresponding to neutron star models with $\rho_{\rm c}\le 2\rho_0$, does not intersect with the region shown in Fig. \ref{fig:MR-sigma}. Thus, to narrow down the allowed region for the neutron star models corresponding to GRB 200415A, we will make a further assumption so that the expected mass and radius could somehow intersect with the region shown in Fig. \ref{fig:MR-sigma}. 

As seen in Fig. \ref{fig:MR-QPO}, the masses of the neutron star models with $\rho_{\rm c}\le 2\rho_0$ are quite small and they do not overlap with the constraint mass vs radius area found in  Fig. \ref{fig:MR-sigma}. Still, the neutron star radius hardly changes at all in the domain of $M\simeq 0.5-1.5 M_\odot$ for any EOS. Based on this observation we can extrapolate our mass vs radius domain as constrained by the vertical dashed lines corresponding to $R=11.73$, $12.41$, $13.03$, and $13.23$ km for $\eta=70.6$, $90.5$, $111.5$, and $118.5$ MeV. In this way, we create an overlapping of the constrained region based on the information extracted from  GRB 200415A, as shown in  Fig. \ref{fig:MR-sigma}, with the neutron star models based on the allowed values of $\eta$ ($\eta_{\rm QPO}$).

%%%%%%%%%%%%%%%%%%%%%%%%%%%%%%%%%%%
% Figure 7
%%%%%%%%%%%%%%%%%%%%%%%%%%%%%%%%%%%
%%%%%%%%%%%%%%%%%%%%%%%%%%%%%%%%%%% 

\begin{figure*}
\begin{center}
\includegraphics[scale=1.1]{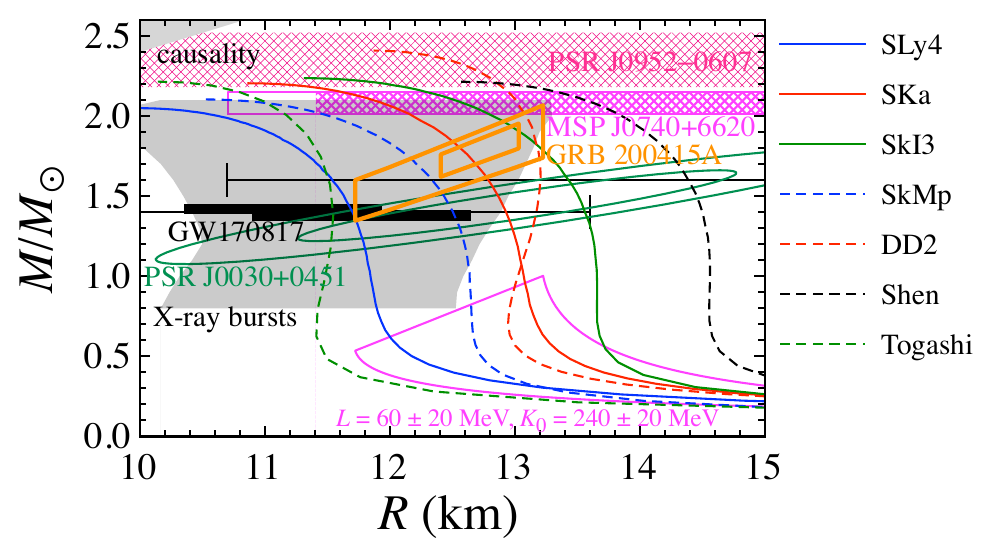} 
\end{center}
\caption{%%
The neutron star mass and radius constrains in this study from GRB 200415A (double-parallelogram) are shown together with other observational and experimental constraints (see the text for details). For reference, we also plot the neutron star mass and radius theoretically constructed with several realistic EOSs.
}%%
\label{fig:MR}
\end{figure*}

Finally, the allowed mass vs radius area for the neutron star models constrained by the observed QPOs in GRB 200415A is compared to the other constraints on the neutron star mass and radius obtained from the astronomical and experimental observations. Actually, in Fig.~\ref{fig:MR} we define, with the two parallelograms, the allowed region for the neutron star model that matches the information extracted from GRB 200415A.  The outer (inner) parallelogram corresponds to the expected region assuming $\eta = 70.6-118.5$ MeV ($\eta_{\rm QPO} = 90.5-111.5$ MeV).
Meanwhile, in the same figure,  we plot the constraints on the neutron star mass and radius set by
(i)   NICER observations for PSR J0030+0451 and MSP J0740+6620 \citep{Riley19,Miller19,Riley21,Miller21}; 
(ii)  observations of x-ray bursts \citep{SLB2013}; and 
(iii)  gravitational wave observations and constraints set by GW170817. 
Especially, for GW170817, we draw the conservative constraint set by the tidal deformability, predicting that the radius of a neutron star with $M=1.4M_\odot$ should be $R_{1.4}\le 13.6$km \citep{Annala18} and that with $M=1.6M_\odot$ should be $R_{1.6}\ge 10.7$km \citep{Bauswein17}, together with more stringent constraints obtained from the combination of multimessenger observations and nuclear theory, i.e., $R_{1.4}= 11.0^{+0.9}_{-0.6}$ km \citep{Capano2020} and $R_{1.4}= 11.75^{+0.86}_{-0.81}$ km \citep{D20}. 

Recently, a new upper neutron star mass limit was set by PSR J0952-0607,  the so-called ``black widow'', 
%is observed in the disk of the Milky Way by Keck-telescope, 
that is $M=2.35\pm 0.17M_\odot$ \citep{Romani22} which is also shown in Fig. \ref{fig:MR}. In the high mass region we also drew the limits set by causality \citep{Lattimer12}.
Finally, the low-mass region enclosed by the purple solid line 
%on the right-bottom side of the figure represents  low mass neutron star models drawn for  
is based on the constraints set for $L=60\pm 20$ and $K_0=240\pm 20$ MeV, as in  Fig. \ref{fig:MR-QPO}. 
For reference, in Fig.~\ref{fig:MR}, we also show the neutron star models constructed using realistic EOSs. Among the EOSs adopted here, the Shen EOS has already been excluded from the gravitational wave observation in GW170817, we keep it here as a reference since it is one of the standard EOSs in astrophysics.

%%%%%%%%%%%%%%%%%%%%%%%%%%%%%%%%%%%
% Figure 8
%%%%%%%%%%%%%%%%%%%%%%%%%%%%%%%%%%%
%%%%%%%%%%%%%%%%%%%%%%%%%%%%%%%%%%%

%%%%%%%%%%%%%%%%%%%%%%%%%%%%%%%%%%%%%%%%%%%%%%%%
\section{Conclusion}
\label{sec:IV}
%%%%%%%%%%%%%%%%%%%%%%%%%%%%%%%%%%%%%%%%%%%%%%%%

QPOs observed in the magnetar giant flares can be an extremely useful tool for extracting neutron star properties. 
In this study, we examine the possibility for identifying the QPOs of the recently observed GRB 200415A as overtones of crustal torsional oscillations. As a result, we find that the observed QPO frequencies can be identified as the 1st, 2nd, 4th, and 10th overtones  with specific values of 
$\varsigma=(K_0^4 L^5)^{1/9}$. 

%Actually, $\varsigma$ is a combination of the incompressibility ($K_0$) and density-dependence of nuclear symmetry energy ($L$) and depends on the stellar mass and radius.
Then, by comparing the resultant value of $\varsigma$ for a variety of neutron star models to the appropriate range of $\varsigma$ restricted by the values of $L$ and $K_0$ determined by terrestrial experiments (or even the value of $L$ constrained from the previous giant flare observations), we show the possible range of values for the mass and radius of GRB 200415A.
Furthermore, assuming that the neutron star radius is almost the same as that of the neutron star model, whose central density is twice the saturation density, we derived a more stringent constraint for the neutron star mass and radius of GRB 200415A. 
The outcome of our analysis is in good agreement, with the constraints set by other observations in the gravitational and electromagnetic spectrum (Fig. \ref{fig:MR}).

Since only high frequencies have been observed in GRB 200415A, we only consider the identification of them with the overtones of crustal torsional oscillations. But, if lower frequencies will be further observed in the future from a similar object (or in the case of a repeated burst from the same object), one may try to identify them with the fundamental oscillations in the same framework. In such a way, one could check whether the origin of the QPOs is the crustal torsional oscillations.    

In this study, we worked under the assumption that the strength of the magnetic field in GRB 200415A is $\lsim 10^{15}$G, so that torsional oscillations have a short damping time, but the frequencies are still close to the values of pure crustal oscillations. If it turns out that the magnetic field strength in GRB 200415A was higher than $\sim 10^{15}$G, a revised analysis will be necessary, due to a significant effect on the frequency spectrum of torsional oscillations. We emphasize that the alternative model discussed in \cite{CT21} is also a viable solution.

%\newpage
%%%%%%%%%%%%%%%%%%%%%%%%%%%%%%%%%%%%%%%%%%%%%%%%
%\acknowledgments
%%%%%%%%%%%%%%%%%%%%%%%%%%%%%%%%%%%%%%%%%%%%%%%%
\section*{Acknowledgements}
This work is supported in part by Japan Society for the Promotion of Science (JSPS) KAKENHI Grant Numbers 
JP19KK0354 and  % International (A) by Sotani
JP21H01088,  % Kiban(B) by Sotani
and by Pioneering Program of RIKEN for Evolution of Matter in the Universe (r-EMU).

%%%%%%%%%%%%%%%%%%%%%%%%%%%%%%%%%%%%%%%%%%%%%%%%
%\section*{Data availability}
%The data underlying this article will be shared on reasonable request to the corresponding author.
%%%%%%%%%%%%%%%%%%%%%%%%%%%%%%%%%%%%%%%%%%%%%%%%

%\bibliographystyle{mnras}
%\bibliography{references}
%%%%%%%%%%%%%%%%%%%%%%%%%%%%%%%%%%%%%%%%%%%%%%%%

%\appendix 
\begin{appendix}
%%%%%%%%%%%%%%%%%%%%%%%%%%%%%%%%%%%%%%%%%%%%%%%%
\section{Magnetic effects on the crustal torsional oscillations}   % Appendix A
\label{sec:appendix_1}
%%%%%%%%%%%%%%%%%%%%%%%%%%%%%%%%%%%%%%%%%%%%%%%%

In this study, we worked under the assumption that the strength of the magnetic field in GRB 200415A is $\lsim 10^{15}$G, so that torsional oscillations have a short damping time, but the frequencies are still close to the values of pure crustal oscillations. For $B\geq 10^{15}$ G, there is a strong shift in the torsional oscillation frequencies, and the damping due to the ``continuous'' spectrum is even stronger, if one considers particular geometries for the magnetic field \citep{2007MNRAS.377..159L, CBK2009, vHL2011, ColKok2011, Gabler2012, Gabler18}. For mixed poloidal-toroidal fields, the magnetoelastic oscillation spectrum becomes discrete \citep{2012MNRAS.423..811C}.

% The signal reported in \cite{CT21}, seems to be weaker  than that observed in SGR 1806-20, which was estimated to be $\leq (2-4) \times 10^{15}$~G \citep{ColKok2011, Gabler18}. For smaller magnetic field strengths, i.e., $B\leq 10^{15}$ G, the presence of the magnetic field can hardly be imprinted in the oscillation spectra.  For $B\geq 10^{15}$ G  there is shifting in the torsional oscillation frequencies, and a ``continuous'' spectrum appears if one considers restricted geometries for the magnetic field \citep{2007MNRAS.377..159L, CBK2009, vHL2011, ColKok2011, Gabler2012, Gabler18} while for mixed poloidal-toroidal fields the magnetoelastic oscillations spectrum is becoming discrete \citep{2012MNRAS.423..811C}.

The effect of the magnetic field on pure torsional oscillations has been studied in \cite{2007MNRAS.375..261S}. There it was shown for a variety of neutron star models and EOS that the shift in the torsional oscillations frequencies due to magnetic field obeys the following formula
%How the frequencies of the crustal torsional oscillations would be modified as the field strength increases, has been studied and can be expressed as
\begin{equation}
    \frac{{}_\ell f_n}{{}_\ell f^{(0)}_n}\approx \left[1  + {}_\ell\alpha_n \left(\frac{B}{B_\mu}\right)^2\right]^{1/2}, \label{eq:modificaiton}
\end{equation}
where ${}_\ell f^{(0)}_n$ is the torsional mode frequency of the $n$-th overtone of a non-magnetized neutron star, while ${}_\ell f_n$ is the frequency of the equivalent magnetized one with the same parameters ($M$, $R$, EOS). $B$ is the strength of the surface  magnetic field normalized by  $B_\mu=4\times 10^{15} \, {\rm G}$. 
This was the case also in the 2D linear simulations in \citep{ColKok2011, 2012MNRAS.423..811C} and it was almost confirmed  in the most recent  2D nonlinear studies  with a poloidal magnetic field \citep{Gabler18}, where the oscillation frequencies are named as \emph{magnetically modified torsional modes}. 

\begin{figure*}
\begin{center}
\includegraphics[scale=0.6]{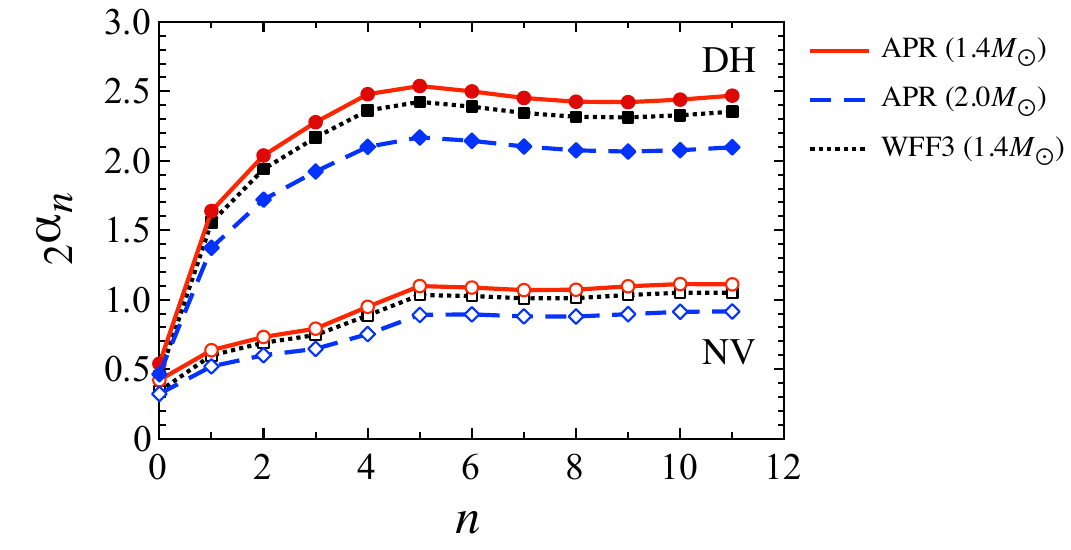} 
\end{center}
\caption{%% 
The coefficients of  ${}_2 a_n$ in Eq. (\ref{eq:modificaiton})  for the neutron star models constructed with two different crust EOSs, NV \citep{NV1973} and DH \citep{DH2001}. For the interior (core) we have used the APR and WFF3 EOS.  The open  and filled marks correspond to  results with NV and DH EOS, while the solid, dashed, and dotted lines correspond to the results for the $1.4M_\odot$, $2.0M_\odot$ neutron star models for the APR EOS, and $1.4M_\odot$ neutron star models for the WFF3 EOS.
}%%
\label{fig:APR14-a}
\end{figure*}

In the previous studies \citep{2007MNRAS.375..261S}, the coefficients ${}_\ell \alpha_n$, have been calculated only for  $n=0, \,1$ and their values vary from 0.3 to 0.5 for EOS NV \citep{NV1973} and 0.4 to 1.5 for EOS DH \citep{DH2001}. 
Here we calculated  ${}_2 \alpha_n$ for larger values of $n$ as it is shown in Figure \ref{fig:APR14-a}. The values seem to reach a maximum value of about ${}_2 \alpha_n \approx 2-2.5$ for EOS DH  and ${}_2 \alpha_n \approx 0.8-1.1$  for EOS NV.
 Thus the deviation of the magnetized neutron star frequencies from those of the non-magnetized ones are $\lsim 3.4\%$ for the EOS NV and $\lsim 7.5\%$ for the EOS DH, if we assume $B \sim 10^{15}$ G. These values are still within the limits of uncertainty  ($\sim 10\%$) estimated in \cite{CT21}. 
For values of the magnetic field significantly higher than  $B\sim 10^{15}$ G, our approach would need to be modified. 

Regarding the damping of crustal oscillations due to the Alfv\'en continuum in the core, we recall that the 3.5ms duration of the high-frequency QPOs observed in GRB 200415A is consistent with the example of a timescale of $\sim 5$ms for the damping of a crustal $n=1$ shear mode in a model with a magnetic field strength of $\sim10^{15}$G shown in \cite{Gabler18}, which justifies our assumptions.

%%%%%%%%%%%%%%%%%%%%%%%%%%%%%%%%%%%
% Figure A1
%%%%%%%%%%%%%%%%%%%%%%%%%%%%%%%%%%%
%%%%%%%%%%%%%%%%%%%%%%%%%%%%%%%%%%%

%\appendix
%%%%%%%%%%%%%%%%%%%%%%%%%%%%%%%%%%%%%%%%%%%%%%%%
\section{Can the 836 Hz QPO be the 2nd overtone?}   % Appendix B
\begin{figure}
\begin{center}
\includegraphics[scale=0.6]{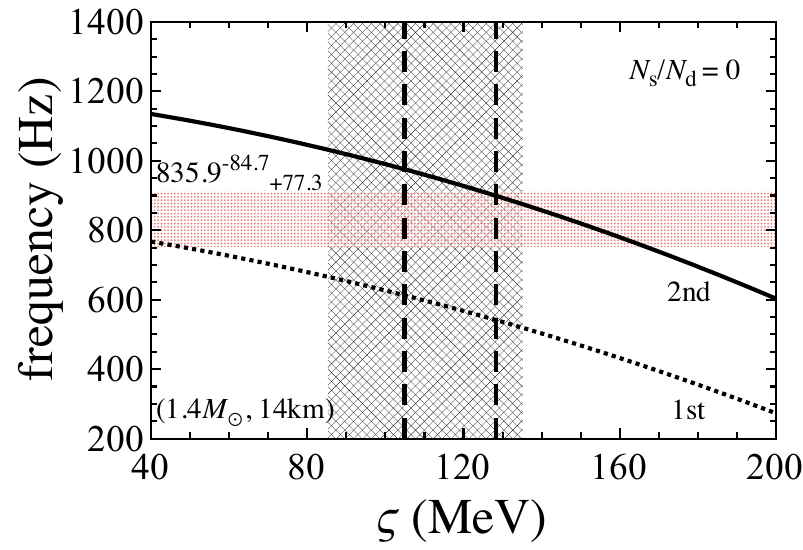} 
\end{center}
\caption{%%
The 1st (dotted line) and 2nd (solid line) overtones  of the $\ell=2$ crustal torsional oscillations for a neutron star model with $(M,R)=(1.4M_\odot,14{\rm km})$ and $N_{\rm s}/N_{\rm d}=0$ is shown as  functions of $\varsigma$. For reference, we plot the lowest observed QPO frequency (red shaded area) in GRB 200415A, i.e., 835.9$^{-84.7}_{+77.3}$ Hz \citep{CT21}. The ranges for $\varsigma$ (dark shaded area) and $\varsigma_{\rm QPO}$ (the area between the dashed lines) are also drawn for comparison. 
}%%
\label{fig:2nd}
\end{figure}

\begin{figure}
\begin{center}
\includegraphics[scale=0.6]{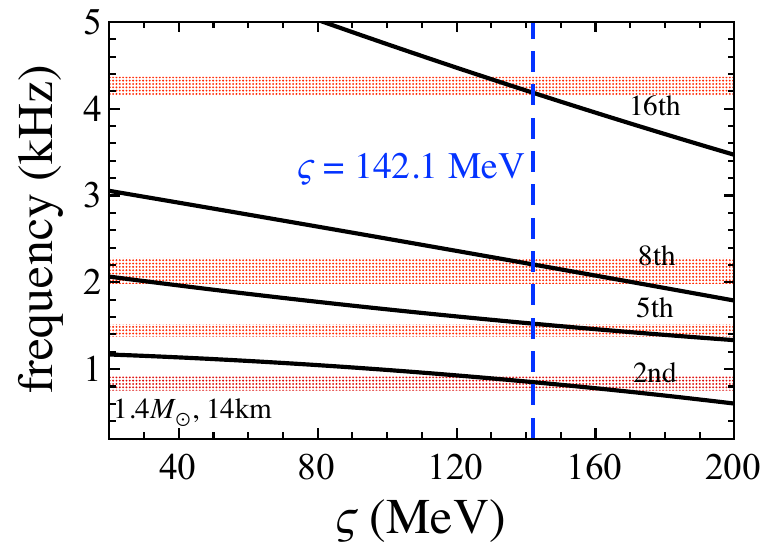} 
\end{center}
\caption{%%
Identification of the QPOs observed in GRB 200415A as crustal torsional oscillation overtones for a neutron star model with $M=1.4M_\odot$ and $R=14$ km, assuming that the lowest QPO in GRB 200415A corresponds to the 2nd overtone. 
}%%
\label{fig:2nd-4QPOs}
\end{figure}

\label{sec:appendix_2}
%%%%%%%%%%%%%%%%%%%%%%%%%%%%%%%%%%%%%%%%%%%%%%%%
In the main part of the article, we identified the lowest QPO frequency at 836 Hz as the 1st overtone. In this appendix, we examine the possibility  of identifying the lowest QPO as the 2nd overtone. 
In Fig. \ref{fig:2nd} we plot the 1st and  2nd overtone of crustal torsional oscillations for a neutron star model with $1.4M_\odot$ and $14$ km as a function of $\varsigma$. In the same figure we draw the observed QPO frequency and the aforementioned values for  $\varsigma$ and $\varsigma_{\rm QPO}$. 
It is apparent that the 836 Hz QPO frequency can be identified as the 2nd overtone. 
In fact, as it is shown in Fig. \ref{fig:2nd-4QPOs}, all four QPOs observed in GRB 200415A can be identified as the 2nd, 5th, 8th, and 16th overtones, if $\varsigma=142.1$ MeV. 
However, as it was mentioned earlier, the overtone frequencies increase with compactness, which implies that the suitable $\varsigma$ values for identifying the observed QPOs should be larger, beyond the accepted values for $\varsigma$. Therefore, for being in agreement with the range $\varsigma=85.3-135.1$ MeV, we would need to consider neutron star models with very small and somehow unphysical compactness.

%%%%%%%%%%%%%%%%%%%%%%%%%%%%%%%%%%%
% Figure A2
%%%%%%%%%%%%%%%%%%%%%%%%%%%%%%%%%%%
%%%%%%%%%%%%%%%%%%%%%%%%%%%%%%%%%%%

%%%%%%%%%%%%%%%%%%%%%%%%%%%%%%%%%%%
% Figure A3
%%%%%%%%%%%%%%%%%%%%%%%%%%%%%%%%%%%
%%%%%%%%%%%%%%%%%%%%%%%%%%%%%%%%%%%

\end{appendix}

\end{document}